\documentclass[
twocolumn,
preprintnumbers,
amsmath,
amssymb,
nofootinbib
]{revtex4}
\usepackage{here}
\usepackage{footnote}
\usepackage{comment,braket}
\usepackage{epsf}
\usepackage{amsmath}
\usepackage{graphics}
\usepackage{amsfonts}
\usepackage{amssymb}
\usepackage{latexsym}
\usepackage{color}
\usepackage{natbib}
\usepackage{graphicx}
\usepackage{hyperref}
\usepackage[svgnames]{xcolor}
\definecolor{phthaloblue}{rgb}{0.0, 0.06, 0.54}
\hypersetup{
    colorlinks=true,
    linkcolor=black,
    citecolor=blue,
    filecolor=blue,
    urlcolor=black,
    }

\preprint{RESCEU-5/20}
\begin{document}

\title{Quantum Black Hole Seismology II: Applications to Astrophysical Black Holes}
\author{Naritaka Oshita$^{1}$\footnote{These authors equally contributed to this work.}}
\author{Daichi Tsuna$^{2,3 \ \ast}$}
\author{Niayesh Afshordi$^{1,4,5}$}
\affiliation{
  $^1$Perimeter Institute, 31 Caroline St., Waterloo, Ontario, N2L 2Y5, Canada
}

\affiliation{
  $^2$Research Center for the Early Universe (RESCEU), the University of Tokyo, Hongo, Tokyo 113-0033, Japan
}
\affiliation{
  $^3$Department of Physics, School of Science, the University of Tokyo, Hongo, Tokyo 113-0033, Japan
}

\affiliation{
  $^4$Department of Physics and Astronomy, University of Waterloo,
200 University Ave W, N2L 3G1, Waterloo, Canada
}

\affiliation{
  $^5$Waterloo Centre for Astrophysics, University of Waterloo, Waterloo, ON, N2L 3G1, Canada
}

\begin{abstract}
With the advent of gravitational wave astronomy, searching for gravitational wave echoes from black holes (BHs) is becoming an interesting probe of their quantum nature near their horizons. Newborn BHs may be strong emitters of echoes, as they accompany large perturbations in the surrounding spacetime upon formation. Utilizing {\it the Quantum Black Hole Seismology} framework \cite{Oshita:2020dox}, we study the expected echoes upon BH formation resulting from neutron star mergers and failed supernovae. For BH remnants from neutron star mergers, we evaluate the consistency of these models with the recent claim on the existence of echoes following the neutron star merger event GW170817. We find that the claimed echoes in GW170817, if real, suggest that overtones contribute a significant amount of energy in the ringdown of the remnant BH. We finally discuss the detectability of echoes from failed supernovae by second and third-generation gravitational wave detectors, and find that current (future) detectors constrain physical reflectivity models for events occurring within a few Mpc (a few $\times$ 10 Mpc). Detecting such echo signals may significantly constrain the maximum mass and equation of state of neutron stars.
\end{abstract}

\maketitle

\section{Introduction}
The recent detection of gravitational waves (GWs) from the merger of binary black holes (BHs), GW150914, has marked the beginning of gravitational wave astronomy \cite{Abbott:2016blz}. GWs not only enable us to investigate the emission and properties of these compact objects but also the compact objects themselves, for instance whether BHs are really ``black" down at their horizons.

GW echoes are proposed as observable smoking guns of quantum effects near BH horizons \cite{Kawai:2015uya,Oshita:2018fqu,Cardoso:2019apo,Oshita:2019sat,Ho:2019qiu,Kawai:2020rmt}, or for alternatives to BHs, often referred to as exotic compact objects \cite{Mazur:2001fv,Schunck:2003kk,PhysRevLett.61.1446,Mathur:2005zp,Cardoso:2016rao,Cardoso:2016oxy}. A partially reflective (rather than completely absorptive) boundary at a microscopic distance (e.g., Planck length) from the would-be horizon would trap GWs between the horizon and the angular momentum barrier, yielding delayed late-time signals (i.e. GW echoes) after the ringdown phase of BH merger/formation. The robust outcome would be a repeating signal of interval (e.g., \cite{Abedi:2018npz})
\begin{equation}
\Delta t_{\rm echo} \sim \frac{4GM}{c^3}\ln\left(\frac{\gamma M}{M_{\rm Pl}} \right) \left[\frac{1}{\sqrt{1-\bar{a}^2}}+1\right],
\label{eq:fecho}
\end{equation}
where $M$ and $\bar{a}$ are the mass and (dimensionless) spin of the BH, $G$ is the gravitational constant, $c$ is the speed of light, and $M_{\rm Pl}$ is the Planck mass. Here, the reflection is assumed to happen at a proper distance of $\gamma \times$ Planck length from the would-be horizon.
While one may expect $\gamma \sim 1$ for Planck-scale modifications, it could also take values $\ll 1$ ($\gg 1$) if the process that is responsible for echoes takes place at super-Planckian (sub-Planckian) energies. 

 Let us assume the case of a BH formed from a collapse of a massive neutron star (NS). For a BH of mass $2.5\ M_\odot$ and spin $0.7$, $\Delta t_{\rm echo}$ is $\sim 10$ ms. The lowest harmonic of this echo signal is thus $\sim 100$ Hz, which is right in the frequency band where the ground-based GW detectors are the most sensitive.
Motivated by this, search for echo signals from a possible BH remnant of the NS merger event GW170817 \cite{TheLIGOScientific:2017qsa} was recently done in \cite{Abedi:2018npz}, claiming a tentative ($4.2 \sigma$) detection of GW echoes $\sim 1$ second after the time of merger. The fundamental frequency was at $72$ Hz, which is consistent with a high-spin BH remnant of a NS merger, with $\bar{a} = 0.84-0.87$ (for $\gamma$ of order $1$). Interestingly, this spin range is consistent with expectations from numerical simulations of binary neutron star (BNS) mergers \cite{Kastaun:2013mv}, while the time of GW echo peak emission coincides with the BH collapse time $t_{\rm coll} = 0.98^{+0.31}_{-0.26}$ sec, inferred from electromagnetic observations \cite{Gill:2019bvq}. 

The existence of echoes is still a matter of debate, as other independent searches concluded in a lower significance \cite{Conklin:2017lwb}, or with a negative result \cite{Tsang:2019zra,Salemi:2019uea} (albeit, using a method with higher detection threshold than the claimed signal). As independent observations are yet to converge on a consensus (although see \cite{Abedi:2020sgg} for one possible explanation of these discrepancies), it may be important to test whether the theoretical modellng of echo signals can explain their tentative detection. Moreover, theoretical considerations can provide realistic physical targets,  and forecast the feasibility of observing echoes in the future. 

Theoretical modeling of echo signals has been intensively studied over the past few years since the detection of GWs (e.g., \cite{Cardoso:2016oxy,Volkel:2017kfj,Volkel:2018hwb, Wang:2018gin, Testa:2018bzd, Oshita:2018fqu, Oshita:2019sat, Wang:2019rcf, Maggio:2019zyv,Conklin:2019fcs}). One toy model widely used is that the reflectivity near the BH horizon is independent of frequency, what we call ``the constant reflectivity'' (or CR) model. More recently, a physically motivated reflectivity model, the Boltzmann reflectivity (BR), was introduced that exponentially depends on the ratio of the frequency to BH's Hawking temperature \cite{Oshita:2018fqu, Oshita:2019sat}.

The formulations for modeling echo signals from realistic spinning quantum BHs by using the Chandrasekhar-Detweiler (CD) equation \cite{Chandrasekhar:1976zz,Detweiler:1977gy}, was introduced under the title of {\it Quantum Black Hole Seismology} in \cite{Oshita:2020dox} (hereafter Paper I), which calculated the spectrum of echoes under both CR and BR models. In this work, we apply the modeling to BH remnants from two major astrophysical scenarios for BH formation, NS mergers and failed supernovae.

In the next section, we briefly summarize Paper I. In Sec. \ref{sec:application}, our echo model is applied to NS mergers and failed supernovae. First we focus on echoes from BH remnants of NS mergers, and discuss the consistency between the tentative detection of echo signals in GW170817 and the BR/CR models. Next we consider BHs from failed supernovae, and discuss the detectability of echo signals in future GW observations. The final section is devoted to conclusions. 

\section{Overview of Paper I \cite{Oshita:2020dox}}
The reflectivity of GWs at the horizon is determined by the quantum nature of the BHs near the horizon. Paper I assumed the two aforementioned models of the reflectivity expressed as
\begin{align}
{\mathcal R} =
\begin{cases}
R_c e^{i \delta_{\text{wall}}} & \text{CR model},\\
\exp\left({- \displaystyle \frac{\hbar|\tilde{\omega}|}{2 k_B T_{\text{QH}}}} +i \delta_{\text{wall}} \right) & \text{BR model},
\end{cases}
\label{ref_con_bol}
\end{align}
where $\tilde{\omega} \equiv \omega -m \Omega_{\rm H}$ with $ \Omega_H \equiv \frac{\bar{a} c^3}{2GM (1+\sqrt{1-\bar{a}^2})}$ and $m$ is the azimuthal harmonic number, $R_c$ is a constant reflectivity, and $T_{\rm QH}$ is the effective temperature of the quantum horizon. Different independent derivations for BR model was provided in \cite{Oshita:2018fqu,Oshita:2019sat} . For example, if the dispersion relation is modified near the BH horizon as
\begin{equation}
\tilde{\Omega}^2 = (\tilde{K}c)^2 +i  \gamma  \left(\frac{\hbar\tilde{\Omega}}{M_{\rm Pl} c^2}\right) (\tilde{K}c)^2 -\hbar^2 C_d^2 \frac{(\tilde{K}c)^4}{(M_{\rm Pl} c^2)^2}, 
\label{dispersion}
\end{equation}
where $\tilde{\Omega}$ and $\tilde{K}$ are proper frequency and proper wave number of GWs, respectively (while $\gamma$ and $C_d$ are dimensionless constants), then  $T_{\rm QH}$ in Equation (\ref{ref_con_bol}) takes the form
\begin{align}
T_{\rm QH} \simeq
\begin{cases}
\frac{\pi (1+4 C_d^2/\gamma^2)}{\sqrt{2 + 4 C_d^2 /\gamma^2}} T_{\rm H} \ \ &\text{for} \ \ C_d \gg \gamma,\\
T_{\rm H} \ \ &\text{for} \ \ C_d \ll \gamma,
\end{cases}
\end{align}
where $T_{\rm H}$ is the (classical) Hawking temperature
\begin{equation}
k_B T_{\rm H} = \frac{\hbar c^3}{4\pi GM}\left(\frac{\sqrt{1-\bar{a}^2}}{1+\sqrt{1-\bar{a}^2}}\right).
\end{equation}
The parameter $\delta_{\rm wall}$ appearing in both models is the phase shift due upon reflection. 

Using the reflectivity at the horizon and the angular momentum barrier as boundary conditions, Paper I numerically solved the CD equations in Kerr spacetime to obtain the spectra of the echoes that reach the observer. The initial signal is assumed to be a ringdown signal, with its total energy parameterized as $\epsilon_{\rm rd} Mc^2$. For simplicity we assume a single quasi-normal mode, i.e. assume one mode energetically dominates over all the other modes. Examples of the echo spectra for different reflectivity models, BH spins, and ringdown modes are shown in its Figures 12, 13, and 14. One notable finding in Paper I was that overtones highly excite the low-frequency region, giving echo amplitudes as much as an order of magnitude larger than the fundamental mode.  This is because the overtones give a sharper wave packet in the time domain than the fundamental modes, due to their rapid decays. This results in a broader spectrum in the frequency domain, which can give more spectral power to the lower frequencies.

Paper I also showed that there exists a bound on the reflectivity parameters $R_c$ and $T_{\rm QH}$ in order to avoid the ergoregion instability. We find upper bounds $R_c \lesssim 0.72$ and $T_{\rm QH} \lesssim 2T_{\rm H}$ for BHs at maximal spin $\bar{a}=0.998$\footnote{Teukolsky numerically calculated the amplification factor for the near-extremal case $\bar{a}=0.99999$ and obtained  $1.38$ \cite{1974ApJ...193..443T}. Then Maggio \textit{et al.} pointed out \cite{PhysRevD.99.064007} that the maximum reflectivity should be around $0.64$ to quench the ergoregion instability for any spin.}, while the bounds become looser at smaller spins. The dependence of the upper bounds on spin can be seen in Figures 1 and 2 of Paper I.

\section{Application}
\label{sec:application}
\subsection{Neutron star mergers}
The remnant of the merger of two NSs is considered to have a variety, depending on their masses and equations of state (for a review see e.g. \cite{Bartos:2012vd,Shibata:2015:NR:2904075}). The merger remnants may promptly collapse into a BH within a dynamical timescale (e.g. \cite{Shibata:2006nm,Bauswein:2013jpa}), collapse with a delay due to rotational and thermal pressure support (e.g. \cite{Baumgarte:1999cq,Kastaun:2014fna}), or remain as stable NSs (e.g. \cite{Giacomazzo:2013uua}). Electromagnetic counterparts such as short gamma-ray bursts or kilonovae/macronovae may be observable, which can be used to infer the nature of the remnant. For the BNS merger event GW170817 \cite{TheLIGOScientific:2017qsa}, observations of the subsequent kilonova/macronova emission favor a delayed collapse scenario, although a stable NS remnant is not completely excluded \cite{Margalit:2017dij,Shibata:2017xdx,Shibata:2019ctb,Gill:2019bvq}. A search for GWs from a possible long-lived NS remnant was done \cite{Abbott:2018hgk}, although the upper limit on the emitted GW energy is not very constraining.

Tentative evidence for GW echoes from a BH remnant of GW170817 has been reported by Abedi {\it et al.} \cite{Abedi:2018npz} (for other independent searches regarding this event see \cite{Conklin:2017lwb,Tsang:2019zra}). They reported a significant excess power at  (integer multiples of) $f_{\rm echo} = 72\ (\pm 0.5)$ Hz, peaked around $\sim$1 second after the BNS merger (consistent with the BH collapse time reported in \cite{Gill:2019bvq}). 
\begin{figure*}[t]
    \includegraphics[width=0.9\textwidth]{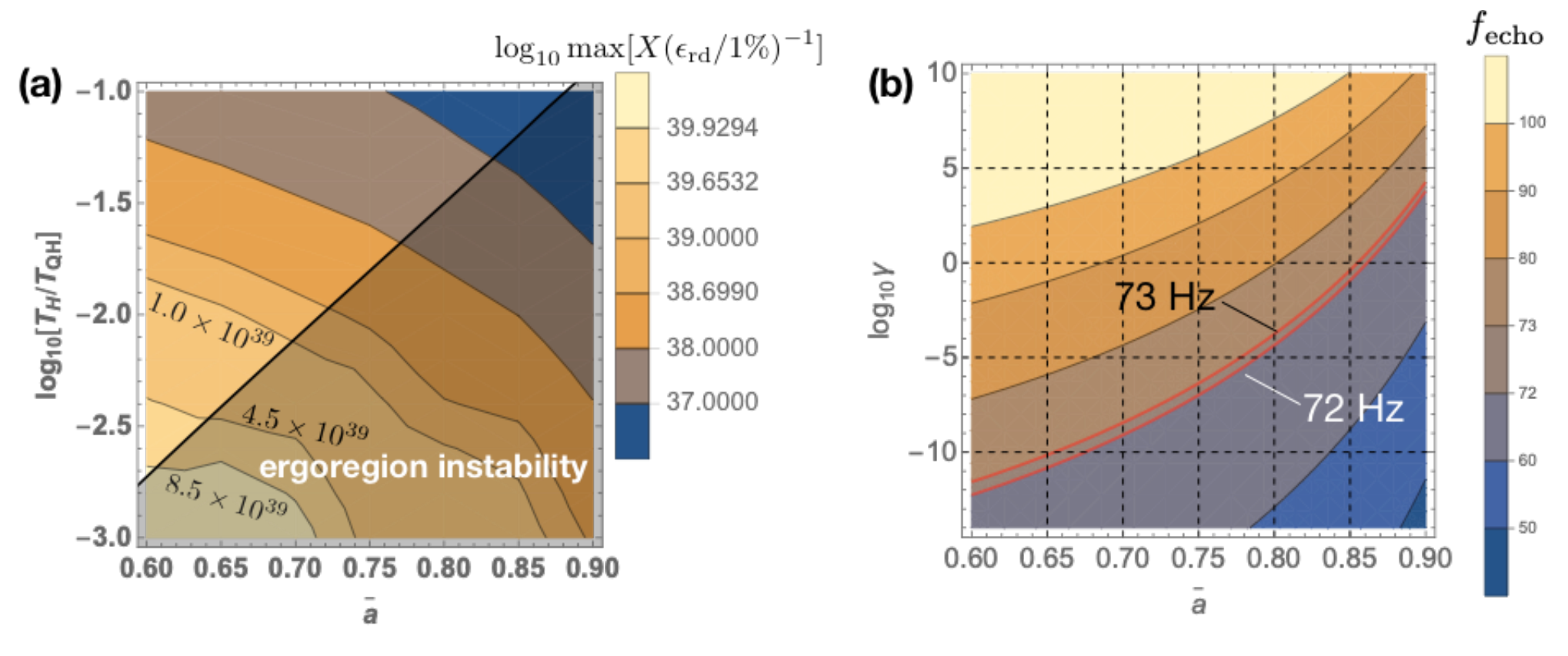}
\caption{(a) Value of the peak of $X(f)$ for the BR model. We take $\gamma = 10^{-10}$ for different values of $T_{\rm H}/T_{\rm QH}$ and the spin of the remnant BH. Gray shaded region is disfavored as it results in an ergoregion instability. Here, we set $\ell = m = 2$ and $n=0$.
(b) Constraints on $\gamma$ and the spin of the remnant BH from the frequency (72 Hz) of the claimed echoes in GW170817. A BH mass of $2.7M_\odot$ is assumed in both plots, and for all the subsequent figures appearing in Sec. III-A.
}
\label{alpha_a_X_2}
\end{figure*}

In this section, we first investigate the parameter region in the BR and CR models which could be consistent with the echo properties reported in \cite{Abedi:2018npz}. Concretely, we try to reproduce their results $f_{\text{echo}} = 72$ Hz and $X(f_{\text{echo}}) = 6.5\times 10^{39}$ $(\text{strain})^{-2}$, with $X (f)$ defined as
\begin{eqnarray} \displaystyle
X (f) = \sum_{k=1}^{10} &\text{Re}& [H(t-t_{\rm merger}=1\ \text{sec},k f) \nonumber \\
&\times& L^{\ast} (t-t_{\rm merger}=1\ \text{sec},k f)].
\end{eqnarray}
Here, the functions $H$ and $L$ are the Weiner-filtered (not whitened, which would have been dividing by square root of PSD). Advanced LIGO observations, defined using the publicly available data and power spectral densities (PSDs) from the Hanford and Livingston detectors obtained from this data\footnote{https://www.gw-openscience.org/events/GW170817/}:
\begin{align}
&H(t,f) = \text{Spectrogram}\left[\text{IFFT} \left(\frac{\text{FFT} (h_H(t-\delta t))}{\text{Hanford PSD}}\right)\right] \label{eq:spec_H} \\
&L(t,f) = \text{Spectrogram}\left[\text{IFFT}\left(\frac{\text{FFT} (h_L(t))}{\text{Livingston PSD} }\right)\right]. \label{eq:spec_L}
\end{align}
Here FFT and IFFT are Fourier and inverse Fourier transforms, and $\delta t \approx 2.62$ ms is the offset of the event between the two detectors, obtained in \cite{Abedi:2018npz} using the premerger inspiral GW signal. We follow the same methodology as \cite{Abedi:2018npz} to obtain the PSD and calculate $H$, $L$ and $X$, except the assumption that the Fourier-transformed strain FFT$(h_H)$ and FFT$(h_L)$ to be the echo spectrum, meaning we neglect the noise component. This assumption is not too bad when considering only the peak, since the contribution to $X(f)$ from noise is at most $\sim 1\times 10^{39}$ as seen in (e.g., Figure 5 of) \cite{Abedi:2018npz}. This assumption also allows a simple scaling $X \propto \epsilon_{\rm rd}$. 

Since \cite{Abedi:2018npz} conducted FFTs of 1 second data segments to calculate $X(f)$, we take the spectral resolution of the signal to be 1 Hz. The spectrogram function has an ambiguity in the normalization, which we need care when doing comparison using $X$. For the spectrogram function \cite{Abedi:2018npz} has adopted, we find that the correct normalization in our methodology (based on our theoretical echo waveform) to have a value consistent with \cite{Abedi:2018npz} is to divide by $2048$, i.e.
\begin{align}
&H(t,f)|_{\rm theory} = \frac{1}{2048} \frac{\tilde{h}_H^{\textrm{(echo)}}}{\textrm{Hanford PSD}} \\
&L(t,f)|_{\rm theory} = \frac{1}{2048} \frac{\tilde{h}_L^{\textrm{(echo)}}}{\textrm{Livingston PSD}}
\end{align}
where $\tilde{h}_{\rm H,\ L}^{\textrm{(echo)}}$ are the echo spectrum defined afterwards. We tested whether the normalization of $X(f)$ is correct by repeating the analysis of \cite{Abedi:2018npz} but injecting an echo signal of known amplitude. We verified that a consistent value of $X(f)$ is obtained at peak frequency.
\begin{figure*}[t]
    \includegraphics[width=1\textwidth]{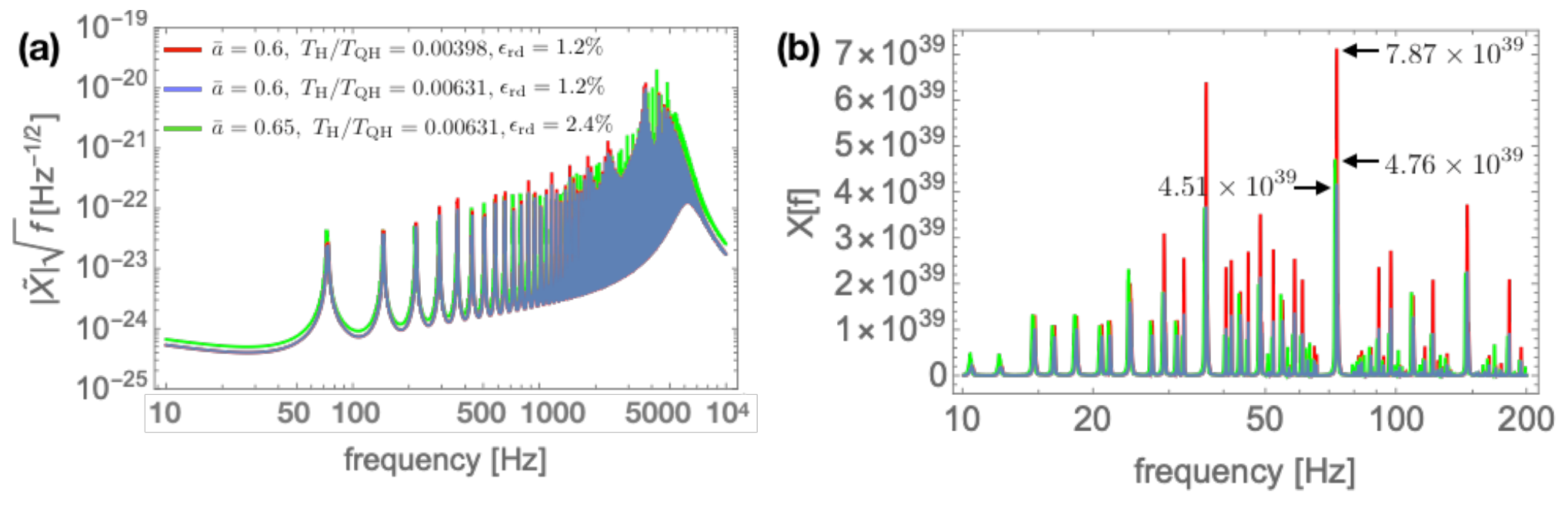}
\caption{Fig. (a) shows the spectra of GW echoes from a remnant BH of $M= 2.7 M_{\odot}$ and $\bar{a}=0.6, \ 0.65$ for $l=m=2$ and $n=0$ in the BR model. The parameter $\gamma$ is set to $10^{-12}$ (red and blue) and $10^{-11}$ (green). Fig. (b) shows the function $X(f)$ with the same parameters.
}
\label{echo_spe_a=0.7_alpha=-2.2}
\end{figure*}

To calculate the theoretical echo signal, we first obtain the ringdown waveform, $h_{H}^{{\rm (rd)}}$ and $h_L^{{\rm (rd)}}$, measured at the detectors in Hanford and Livingston with the absolute values of the respective antenna pattern functions at the location and time of GW170817: ${\cal F} \equiv \sqrt{F_+^2 + F_\times^2} = 0.89$ and $0.75$ \cite{Marra:2019lyc}. The plus and cross modes are modeled as
\begin{align}
h_{+}^{\rm (rd)} &= \frac{M}{D_L} \text{Re} \left[ {\cal A}_{lmn}^{+} e^{-i (\omega_{lmn} t + \phi_{lmn}^{+})} S_{lmn} \right],\\
h_{\times}^{\rm (rd)} &= \frac{M}{D_L} \text{Im} \left[ {\cal A}_{lmn}^{\times} e^{-i (\omega_{lmn} t + \phi_{lmn}^{\times})} S_{lmn} \right],
\end{align}
where $\omega_{lmn}$ is the (complex) QNM associated with $(l,m,n)$ mode, ${\cal A}_{lmn}$ is the amplitude, $S_{lmn}$ is the spheroidal function, and $D_L$ is the luminosity distance assumed to be $40$ Mpc \cite{GBM:2017lvd}.
Following \cite{Flanagan:1997sx} it is assumed that $\phi_{lmn}^+=\phi_{lmn}^{\times}=0$, and the amplitude of plus mode is same as that of cross mode $A^{+}_{lmn}=A^{\times}_{lmn}={\cal A}_{lmn}$ for $m=2$. On the other hand, ${\cal A}^{+}_{lmn} = {\cal A}_{lmn} \neq 0$ and ${\cal A}^{\times}_{lmn} = 0$ for $m=0$ \cite{Gleiser:1998rw,Khanna:2000dg}. In the former case, the observed signal can be modeled as
\begin{align}
&h_{H,L}^{\rm (rd)}(t) = F_+ (\theta_s, \phi_s, \psi_s) h_+^{\rm (rd)} + F_\times (\theta_s, \phi_s, \psi_s) h_\times^{\rm (rd)} \\
\begin{split}
&= \frac{M}{D_L} {\cal A}_{lmn} e^{\text{Im}[\omega_{lmn}] t} |S_{lmn} (\theta, \phi)|\\
&\times \left\{ F_+ \cos{(\text{Re} [\omega_{lmn}] t + \beta)} + F_\times \sin{(\text{Re} [\omega_{lmn}] t + \beta)} \right\}
\end{split}\\
\begin{split}
&= \frac{M}{D_L} {\cal A}_{lmn} e^{\text{Im}[\omega_{lmn}] t} |S_{lmn} (\theta, \phi)|\\
&\times {\cal F} \cos{(\text{Re} [\omega_{lmn}] t + \alpha+\beta)},
\label{time_domain_overserved}
\end{split}
\end{align}
where $\theta_s$, $\phi_s$ are the polar and azimuthal angles in the detector frame, $\psi_s$ is the polarization angle, $\alpha \equiv \cos^{-1} [F_+/{\cal F}]$, and $\beta \equiv \text{arg}(S_{lmn})$. The latter case ($m=0$) is modeled by (\ref{time_domain_overserved}) with $\alpha =0$\footnote{In this manuscript, we assume $|F_+| = |F_{\times}| = {\cal F}/\sqrt{2}$. We also have ${\cal A}_{20n}/{\cal A}_{22n} = \sqrt{2}$ due to the absence of the cross mode for $m=0$.}. The Fourier transform of (\ref{time_domain_overserved}) is 
\begin{equation}
\tilde{h}_{H,L}^{\rm (rd)} (\omega) = \frac{M}{D_L} {\cal A}_{lmn} {\cal F} \left[ e^{i \alpha} S_{lmn} \alpha_+ + e^{- i \alpha} S_{lmn}^{\ast} \alpha_- \right],
\label{spectrum_rd}
\end{equation}
where the functions $\alpha_{\pm} (\omega)$ are defined in Paper I. The phase $\alpha$ should be tuned so that the value of $X$ is maximized, which is equivalent to the tuning of $\delta t$ in \cite{Abedi:2018npz}. The difference of $\alpha$ between the Hanford and Livingston is around $180$ degrees\footnote{The exact number is $164.8$ degrees, and the difference between $180$ and $164.8$ degrees leads to an error of only a few percents in $X$.}, and therefore, we take $\alpha = 0$ or $\pi$ that maximize the function $X$. Multiplying $\tilde{h}_{H,L}^{\rm (rd)}$ by the transfer function introduced in the Paper I we obtain the spectra of GW echoes $\tilde{h}_H^{\textrm{(echo)}}$, $\tilde{h}_L^{\textrm{(echo)}}$, which we plug into equations (\ref{eq:spec_H}) and (\ref{eq:spec_L}) to obtain $X(f)$. The validity of this treatment is discussed in the Appendix \ref{app:h}.

\subsubsection{Fundamental mode, $n=0$}
We first consider the case where the echoes are dominated by the fundamental (or least damped) QNM, which also dominates the late-time ringdown signal.

In the BR model, we have three independent parameters $T_{\rm QH}$, $\delta_{\rm wall}$, and $\gamma$. Here we set the value of $\delta_{\rm wall}$ so that the frequency peaks in spectrum of GW echoes are located at $f = {\rm integers} \times f_{\text{echo}}$, which is consistent with the analysis result in \cite{Abedi:2018npz}. Now we have two parameters $T_{\rm QH}$, $\gamma$ coming for the echo model, and four other parameters for the remnant BH of the BNS merger $\bar{a}$, $M$, $\theta$, and $\epsilon_{\text{rd}}$, where $\theta$ is the viewing angle with respect to the BH spin.

The estimated total mass of the two NSs \cite{TheLIGOScientific:2017qsa,Abbott:2018wiz,LIGOScientific:2019eut} and the subsequent mass loss due to mass ejection and GW emission makes $M$ likely to be around $2.6$--$2.7 M_\odot$. In this subsection, we fix $M$ to be $2.7 M_{\odot}$. We use the estimate on the viewing angle of $\theta \approx 20^\circ$ \cite{Mooley:2018dlz,Hotokezaka:2018dfi} obtained from radio observations of the jet launched after the merger\footnote{This assumes that the BH spin is aligned with the jet launched after the merger, which is natural under the assumption that the BH's rotational energy is used to power the jet.}, which is consistent with the constraint $15^\circ \lesssim \theta \lesssim 40^\circ$ obtained from GWs \cite{TheLIGOScientific:2017qsa,Abbott:2018wiz}. We calculate the spectra of GW echoes with $\ell = m = 2$ and estimate $X(f)$. Using this we constrain the temperature ratio $T_{\rm H}/T_{\rm QH}$ and spin $\bar{a}$ as is shown in Fig. \ref{alpha_a_X_2}-(a). 
One can see that if the remnant BH has a high spin of $\bar{a} \sim 0.85$ (as found in \cite{Abedi:2018npz}, based on the observed $f_{\rm echo}$, and $\gamma \sim 1$), the value of $X$ is at most of order $10^{37}(\epsilon_{\rm rd}/1\%)$. Thus, from energy budget considerations, it is very difficult to reproduce the observed  value of $6\times 10^{39}$. On the other hand, for the case of lower spin of $\bar{a} \lesssim 0.7$ there exists a parameter space that can satisfy the claimed value of $X$. In this case, $T_{\rm H}/T_{\rm QH} < 0.01$ is required, which corresponds to having the reflectivity of $|{\cal R}| \gtrsim 0.94$ for $\omega \lesssim m \Omega_{\rm H}$, and the corresponding value of spin is still in line with numerical relativity simulations; for example simulations modeling GW170817 \cite{Shibata:2017xdx} obtain the spin of the remnant BH as $\bar{a} \sim 0.7$ if it collapsed from a hypermassive NS. This can be even lower if it collapsed from a supramassive NS, due to the longer time available for the NS to transport angular momentum to the remnant torus that is considered to form after merger.
However a low spin requires a small value of $\gamma$ to match the peak frequency at 72 Hz. From Eq. (\ref{eq:fecho}) and fixing the mass as $2.7 M_{\odot}$, one can constrain $\gamma$ and $\bar{a}$ as is shown in Fig.\ref{alpha_a_X_2}-(b). For BH spins of $0.65$ of $\bar{a}$, a small value of $\gamma \sim 10^{-11}$ is required to match the peak frequency. However, it means that the super-Planckian energy around $M_{\text{Pl}} c^2 / \gamma \sim 10^{11} M_{\text{Pl}} c^2$ would be involved at the would-be horizon.

To give a few examples which can be consistent with the tentative detection of GW echoes, three spectra of GW echoes, whose $X (f)$ has a peak of $(4 \sim 7) \times 10^{39}$ at $\sim 73$ Hz, are shown in Fig. \ref{echo_spe_a=0.7_alpha=-2.2}. We thus conclude that the peak frequency and peak value of $X$ in \cite{Abedi:2018npz} can be reproduced with our model within the allowed parameter space. Considering peaks other than the fundamental mode of 72 Hz, such as 36 Hz or 144 Hz, may constrain the parameter space even further.
We note however that extending this low value of $T_{\rm H}/T_{\rm QH}$ to spins higher than $0.7$ leads to the ergoregion instability (Fig. \ref{alpha_a_X_2}(a)). The likely existence of Galactic BHs having much larger spin than this (see e.g. \cite{McClintock:2006xd,Gou:2013dna}) would require $T_{\rm H}/T_{\rm QH}$ to have a nontrivial spin-dependence that can avoid the instability. Another possibility to avoid this apparent tension is by assuming $T_{\rm H}/T_{\rm QH}$ is time dependent instead of spin dependent, i.e. a quantum BH {\it cools down} over time. We do not probe this possibility in detail, as the cooling timescales of these quantum BHs are completely uncertain.

\begin{figure}[t]
    \includegraphics[width=0.5\textwidth]{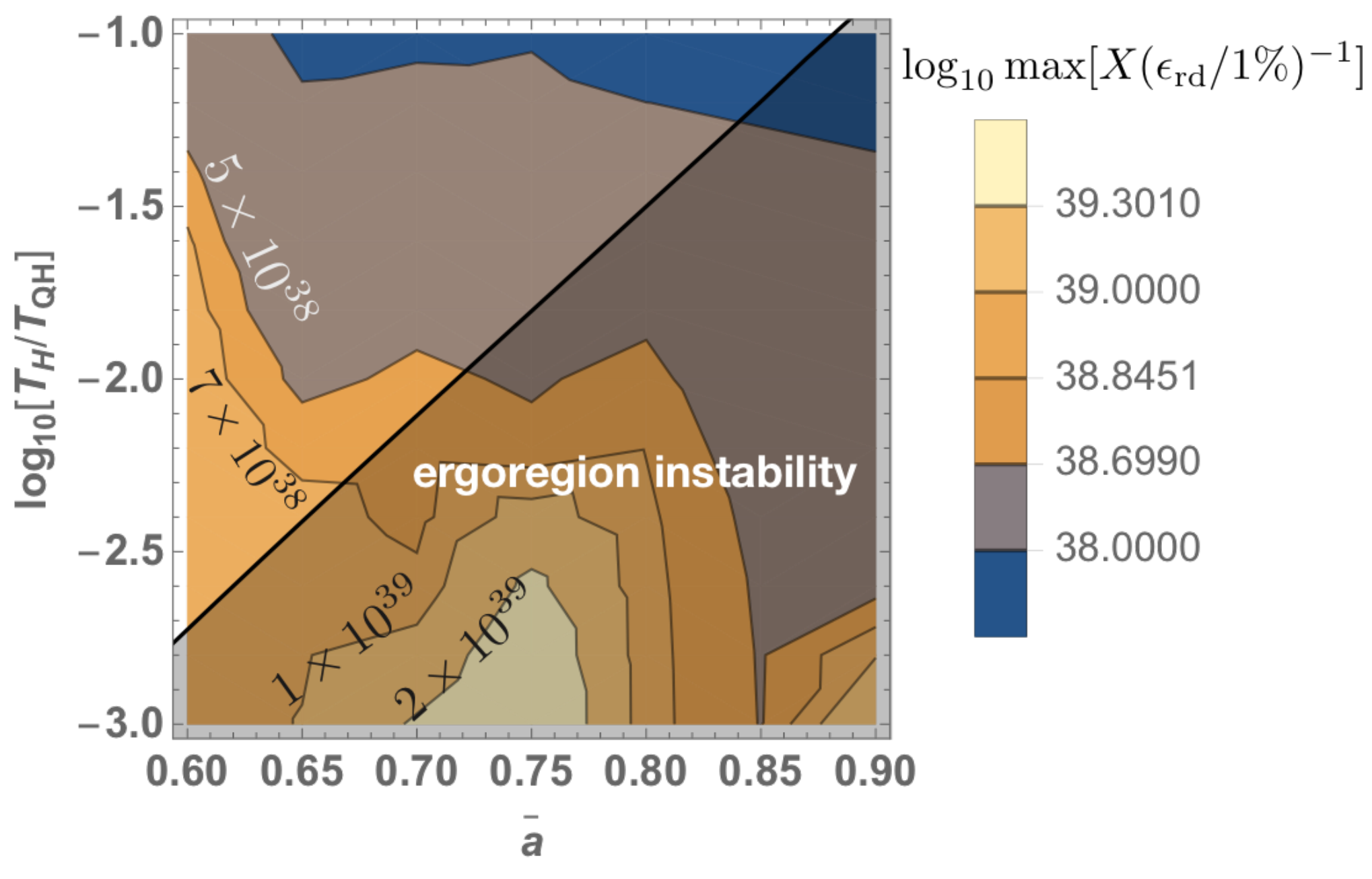}
\caption{Dependence of $X$ obtained in the BR model on $T_{\rm H}/T_{\rm QH}$ and $\bar{a}$ with $\ell=2$, $m=0$, $n=0$, $\gamma = 10^{-10}$, and $\theta =20^\circ$. The grey shaded parameter region leads to the ergoregion instability.
}
\label{contour_alpha_theta_m=0}
\end{figure}

We also consider the possibility where an $\ell =2, m=0$ mode dominated the ringdown GWs emitted after the long-lived NS collapsed into a BH. The dependence of $X$ on $T_{\rm H}/T_{\rm QH}$ and $\bar{a}$ with fixed viewing angle $\theta = 20^\circ$ is shown in Fig. \ref{contour_alpha_theta_m=0}.
\begin{figure*}[t]
    \includegraphics[width=0.9\textwidth]{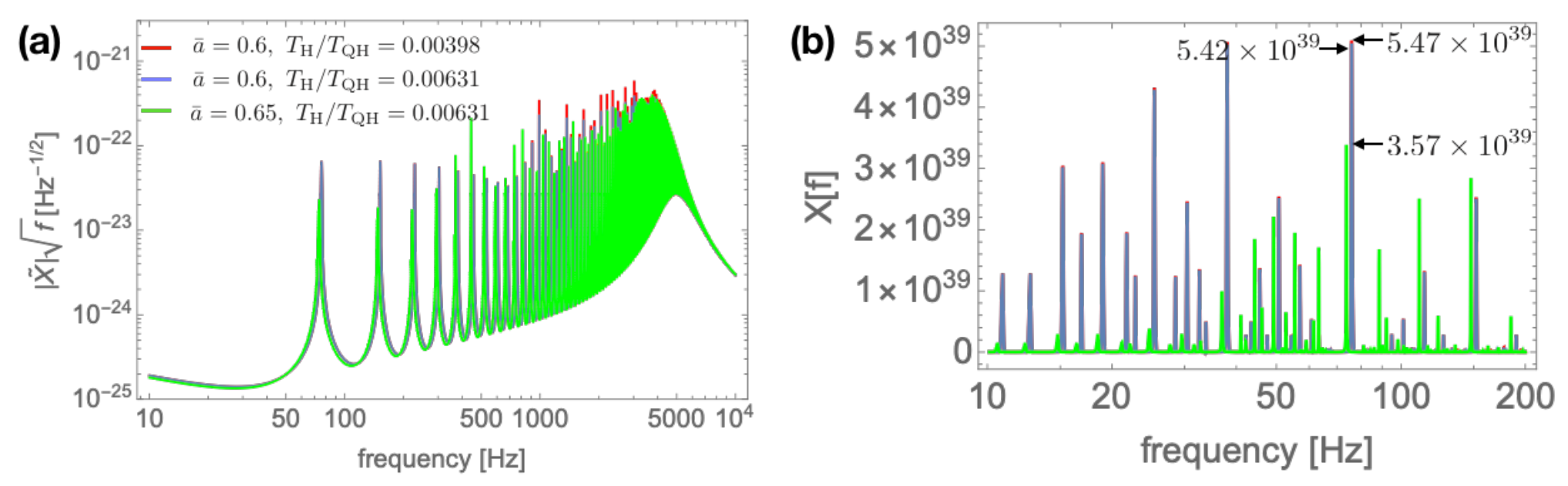}
\caption{Fig. (a) shows the spectra of GW echoes from a remnant BH of $M= 2.7 M_{\odot}$ in the $m=n=0$ case. We use the BR model and the parameter $\gamma$ is set to $10^{-10}$, $\theta = 20^{\circ}$, and $\epsilon_{\rm rd} = 6 \%$. Fig. (b) shows the function $X(f)$ with the same parameters.
}
\label{spectra_m=0_small_alpha}
\end{figure*}
To be consistent with the value of $X = 6.45 \times 10^{39}$, $\epsilon_{\text{rd}}$ should be comparable to or larger than $6 \%$ (Fig. \ref{spectra_m=0_small_alpha}). This requirement is somewhat higher than numerical simulation of BNS mergers (e.g., \cite{Kiuchi:2009jt,Bernuzzi:2015opx,Zappa:2017xba}), which find $1-3\%$ for total emitted post-merger GW energy.
In addition, we see that the $m=0$ case has a large viewing angle dependence as shown in Fig. \ref{echo_X_m=0}-(a), where $T_{\rm H}/T_{\rm QH} = 0.6$ and $M=2.7M_\odot$. However, there is no consistent regions when $T_{\rm H}/T_{\rm QH} = 0.6$. If $T_{\rm QH}$ has a nontrivial dependence on the spin parameter to avoid the ergoregion instability and $T_{\rm H} / T_{\rm QH} \sim 0.002$ around $\bar{a} = 0.6$, one can reproduce $X \sim 6 \times 10^{39}$ (see Fig. \ref{echo_X_m=0}-(b)) with $\epsilon_{\text{rd}} \lesssim 1\%$, for $\theta \gtrsim 40^{\circ}$. However, this is in tension with the viewing angle constrained from the observed inspiral waveform $15^\circ \lesssim \theta \lesssim 40^\circ$ \cite{TheLIGOScientific:2017qsa,Abbott:2018wiz} (see Fig. \ref{echo_X_m=0}-(b)). Nevertheless, if we assume the orbital axis of the two NSs and the resulting BH spin are aligned, it may be possible to avoid the discrepancy by considering a large offset of the NSs' spins with respect to the orbital axis. However, this is very difficult to achieve, since just before merger the angular momenta of the two NSs are likely subdominant compared to the orbital angular momentum \cite{Stone:2012tr}. 

The large reflectivity required to fit $X \sim 6  \times 10^{39}$ (strain)$^{-2}$ also poses a difficulty for the CR model, due to the upper limit on $R_c$ to avoid the ergoregion instability. We find that, for both $m=0$ and $m= 2$ cases, $T_{\rm H}/T_{\rm QH} \lesssim 0.01$ is required, which is approximately equivalent to having $|{\cal R}| \gtrsim 0.94$ in $\omega \lesssim m \Omega_{\rm H}$. Thus the CR model without spin dependence (which requires $R_c < 0.72$ to avoid the ergoregion instability) is also inconsistent with the tentative detection of echoes in \cite{Abedi:2018npz}.
\begin{figure*}[t]
    \includegraphics[width=1\textwidth]{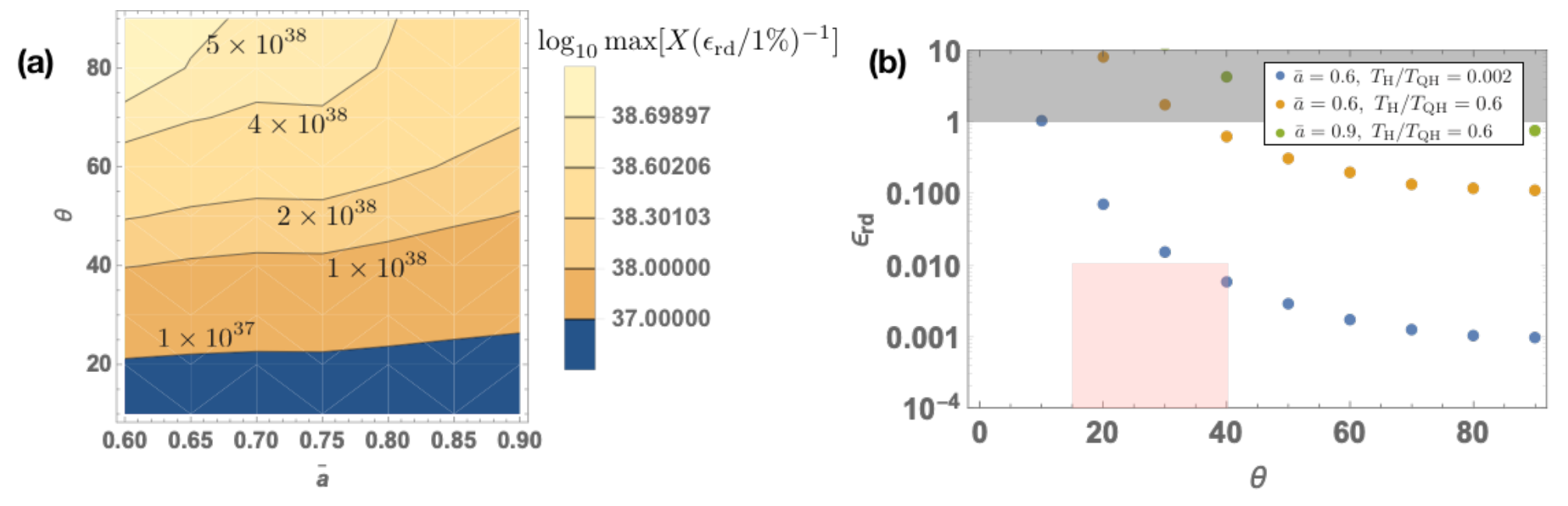}
\caption{(a) Constraints on $\theta$ and spin of the remnant BH of GW170817 from the value of $X(f)$ in \cite{Abedi:2018npz}. Here, the BR model with $T_{\rm H}/T_{\rm QH}=0.6$ is assumed.
(b) Plot of ($\epsilon_{\text{rd}}$, $\theta$) which gives $\text{max}(X(f)) = 6.45 \times 10^{39}$. In both plots, we set $\ell=2$ and $m=n=0$. To be consistent with the observations \cite{Abbott:2018wiz} and total GW emission calculated from numerical simulations (e.g. \cite{Kiuchi:2009jt}), a loose limit of $15 \lesssim \theta \lesssim 40^\circ$ and $\epsilon_{\text{rd}} \lesssim 0.01$ is obtained (red shaded region). $\epsilon_{\text{rd}} >1$ is unphysical (gray shaded region). For both cases, we set $\gamma = 10^{-10}$.
}
\label{echo_X_m=0}
\end{figure*}

\subsubsection{Overtones, $n>0$}
In the previous subsection, we saw that it was difficult to reproduce the claimed echoes following GW170817 if the fundamental ringdown QNM ($n=0$) also dominates the echo signals (leading to ergoregion instability and/or super-Planckian energies). However, the excitation factors of overtones ($n>0$) can be greater than that of the fundamental QNM, and thus they can dominate when the ringdown phase starts. For example, \cite{Giesler:2019uxc} analyzed the waveform of a binary BH merger simulation \cite{Mroue:2013xna} similar with GW150914, and found that the $n=4$ overtone dominates the amplitude near the peak of the merger (even though the relevance of linear theory at this time is debatable). Motivated by this, we next investigate whether the observed echo amplitudes can be reproduced if overtones are energetically dominant in the ringdown. We consider the case where the $n=2$ overtone is dominant, as this results in the loudest echo signal when other parameters are fixed (see Fig. 14 of \cite{Oshita:2020dox}).

We first consider the case where the ringdown consists of highly excited overtone QNMs of $\ell=2$ and $m=0$. The CR model still has a difficulty reproducing the value of $X$, with a required $\epsilon_{\rm rd}\simeq 36\%$. On the other hand, the BR model can reproduce $X \simeq 4 \times 10^{39}$ that is almost consistent with the tentative detection. In addition, the parameters we fixed here do not lead to the unnaturalness of parameters $T_{\rm QH}$ and $\gamma$ mentioned for the above $n=0$ case. We set $\theta = 33^{\circ}$ that is consistent with the constraint on $\theta$ from the GW observation \cite{TheLIGOScientific:2017qsa,Abbott:2018wiz}, $T_{\rm H}/ T_{\rm QH}=0.54$, for which the ergoregion instability is absent (without requiring a nontrivial spin-dependence of $T_{\rm QH}$), and $\gamma = 1$ for which the super-Planckian energy is no longer involved near the would-be horizon. The required value of $\epsilon_{\rm rd}$ is still large; $\epsilon_{\rm rd} = 6.8 \%$ is needed to reproduce $X \simeq 4 \times 10^{39}$.

\begin{figure}[b]
    \includegraphics[width=0.45\textwidth]{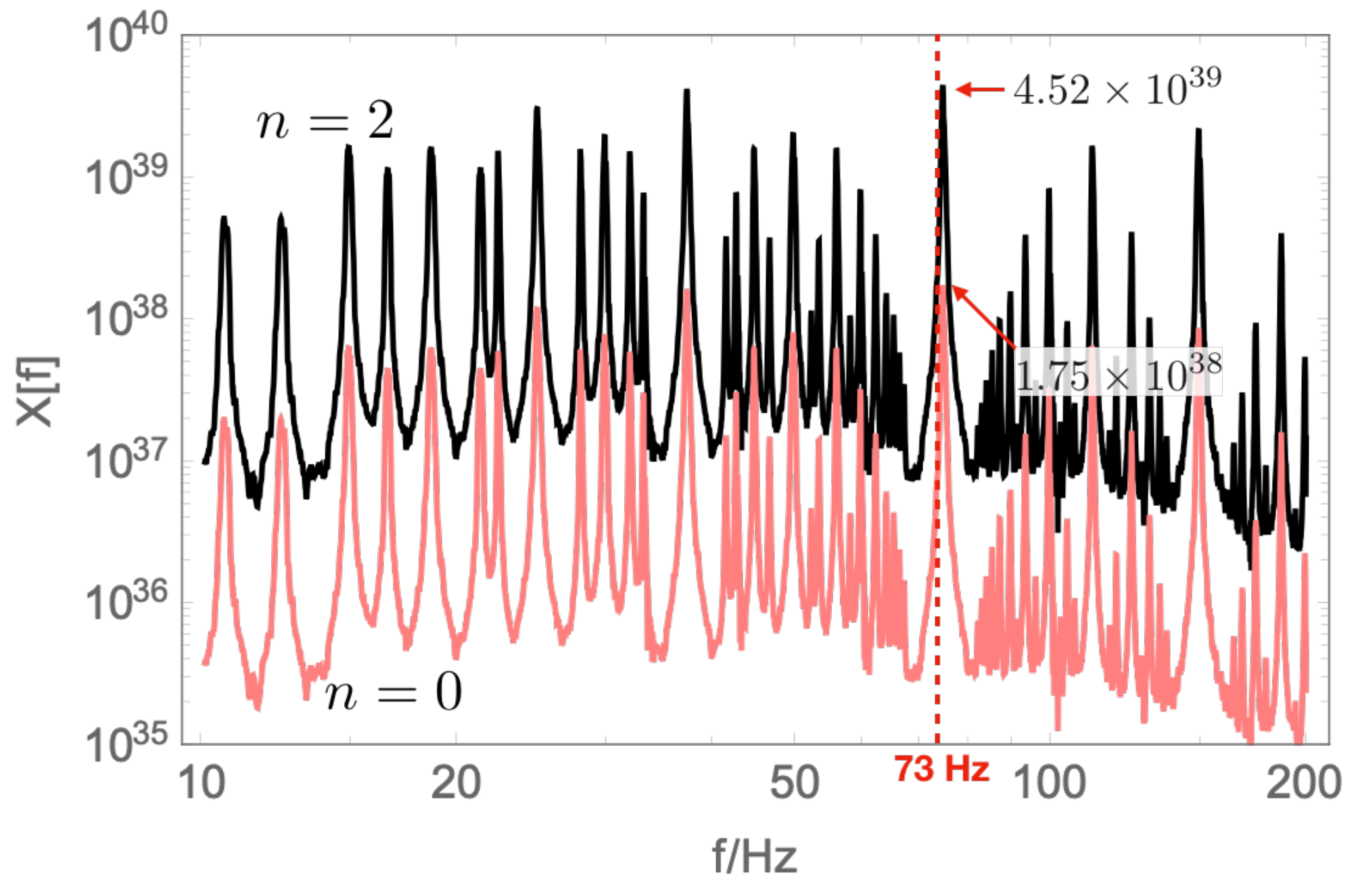}
\caption{Plots of $X(f)$ obtained in the BR model for the overtone QNM with $n=2$ (black) and the least damping QNM (pink). For both cases, we set $\ell =2, \ m=0$, $\bar{a}=0.85$, $\epsilon_{\rm rd} = 0.7 \%$, $\theta = 33^\circ$, $D_L = 40$ Mpc, $T_{\rm H} / T_{\rm QH} = 0.1$, and $\gamma =1$.
}
\label{X_overtone}
\end{figure}

On the other hand, the required value of $\epsilon_{\rm rd}$ can be significantly reduced by a lower $T_H/T_{\rm QH}$. If we set $T_{\rm H} / T_{\rm QH} = 0.1$, one can reproduce $X \simeq 4.5 \times 10^{39}$ with $\epsilon_{\rm rd} = 0.7 \%$ (Fig. \ref{X_overtone}). In this case, a nontrivial spin dependence of $T_{\rm QH}$ is still necessary to avoid the ergoregion instability for near-extremal spins.

\begin{figure}[b]
    \includegraphics[width=0.48\textwidth]{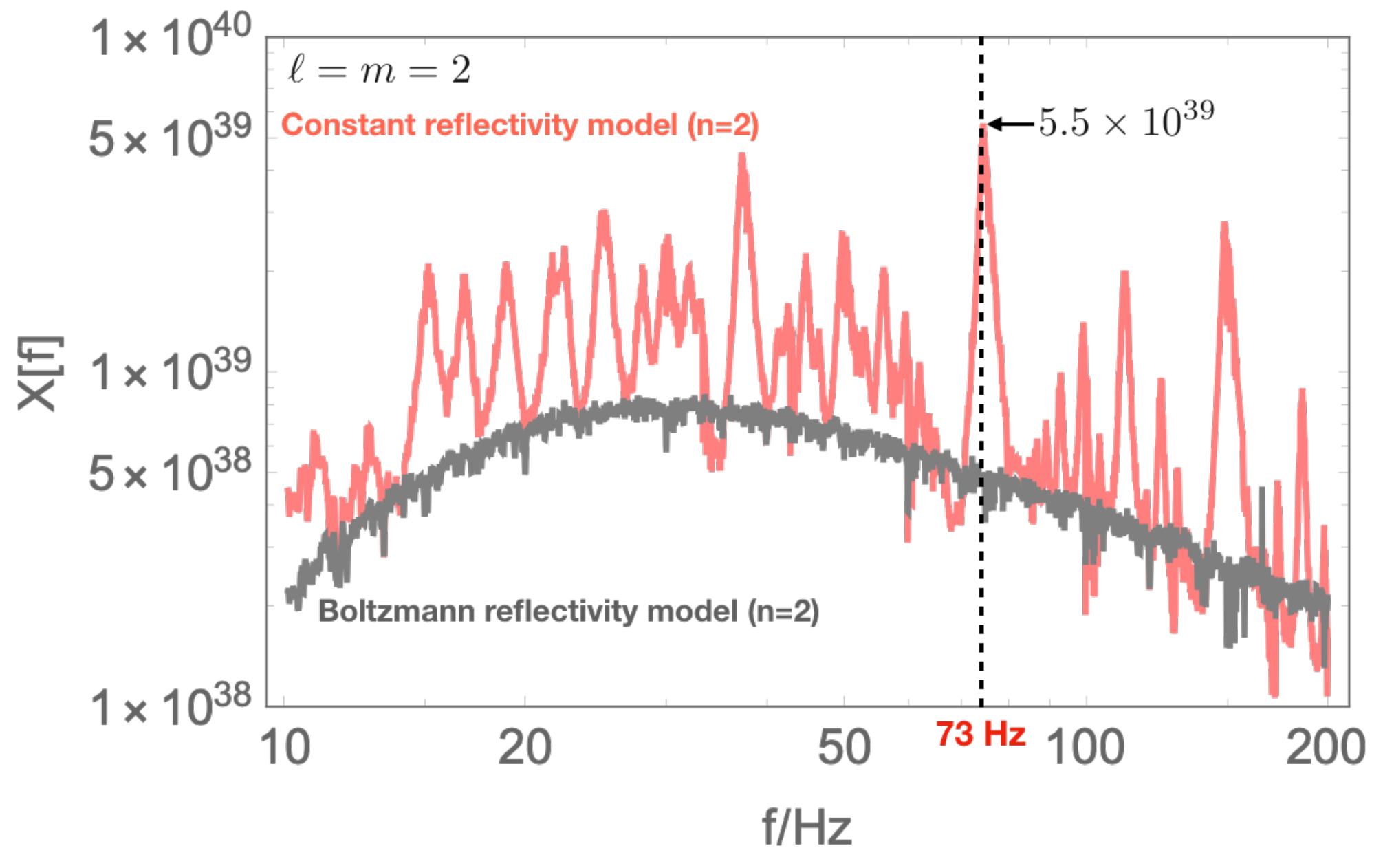}
\caption{Plot of $X(f)$ with $\ell = m = 2$ and a single dominant overtone of $n=2$ in the CR (pink) and BR (gray) models. We set $\bar{a} = 0.85$, $\epsilon_{\rm rd} = 1.2 \%$, $\theta = 20^\circ$, $D_L = 40$ Mpc, $R_c = 0.71$, $T_{\rm H}/T_{\rm QH} = 0.54$, and $\gamma = 1$. 
}
\label{CR_BR_overtone_X}
\end{figure}

\begin{figure}[t]
    \includegraphics[width=0.45\textwidth]{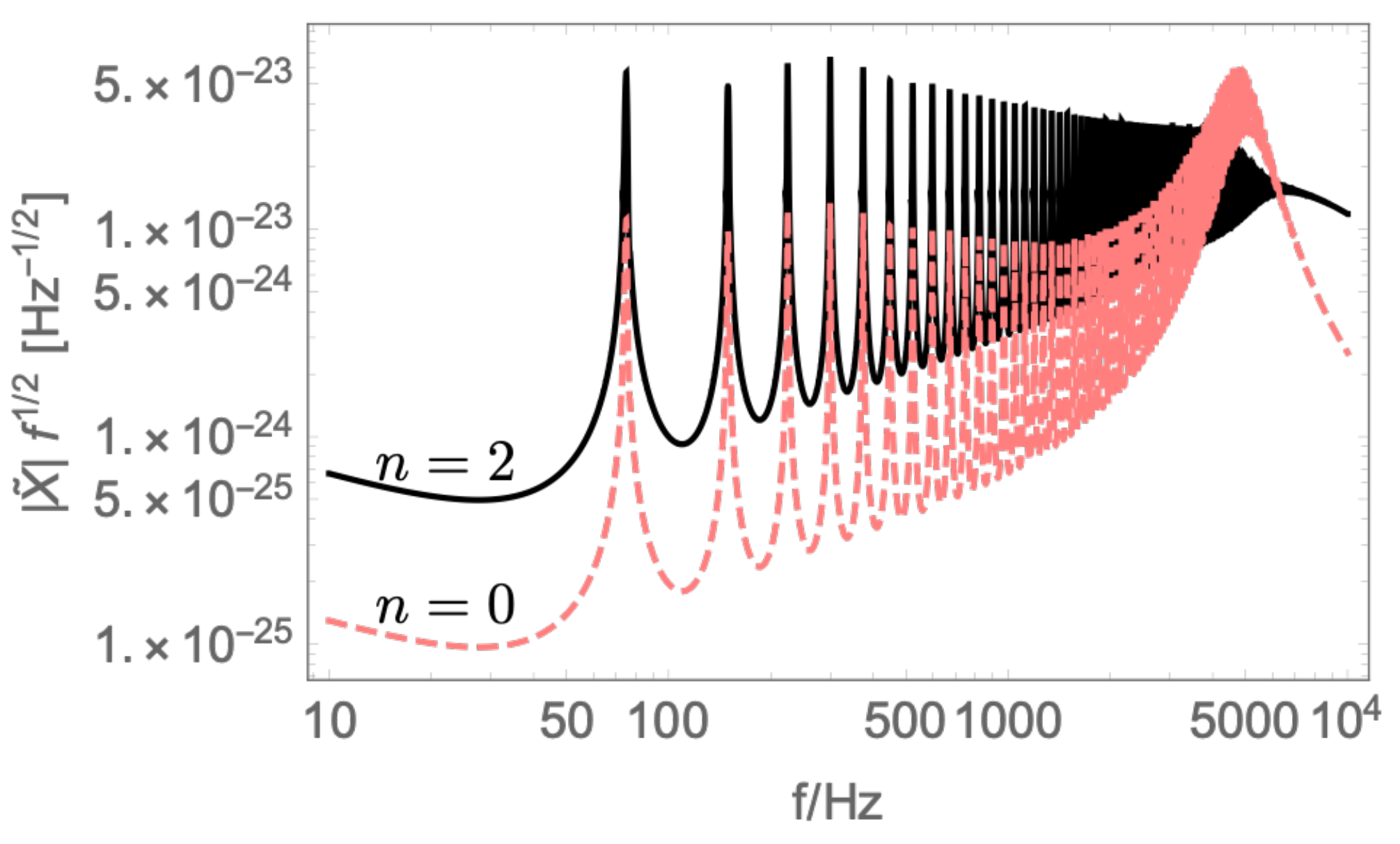}
\caption{Spectra of GW echoes for $n=2$ (black solid) and $n=0$ (pink dashed) in the CR model. The parameters are same as in Fig. \ref{X_overtone}.
}
\label{spectra_overtone}
\end{figure}

If the overtones of $\ell = m = 2$ were dominant in the early ringdown of GW170817, the CR model can be well consistent with the tentative detection whereas the BR model is not (see Fig. \ref{CR_BR_overtone_X}). Here we set $\epsilon_{\rm rd} = 1.2 \%$, $\theta = 20^\circ$, and $R_c = 0.71$. As one can see in Fig. \ref{CR_BR_overtone_X}, the echo peaks are highly suppressed and almost invisible since the reflectivity of the BR model is exponentially suppressed in low-frequency region. For the CR model, the wave packet in frequency domain can be so broad that the echo peaks in low-frequency region are sufficiently enhanced (see Fig. \ref{spectra_overtone}) to be consistent with the tentative detection.

To summarize, the tentative detection of echoes following GW170817, if real, gives stringent constraints for both the CR and BR models (see Table \ref{tab:chart} for an executive summary). In particular, we find that introducing overtones into the ringdown is required. Conversely, echoes can give us information about the overtones in the ringdown, which may otherwise be difficult to probe due to their high frequencies and rapid temporal decays.


\begin{table}
\begin{tabular}{c|c|c}
   & CR model & BR model \\
  \hline
  $m=2, n=0$ & No & No\\
  $m=0, n=0$ & No & No\\
  $m=2, n=2$ & Yes\footnote{If $\epsilon_{\rm rd}\sim$ 1 \%, which is comparable to the {\it total} GW energy expected from NS mergers, using current simulations (c.f. \cite{Kiuchi:2009jt,Bernuzzi:2015opx,Zappa:2017xba}).} & No \\
  $m=0, n=2$ & No
  & Yes\footnote{Fine tuned spin-dependent horizon temperature $T_{\rm QH}$ is required.}\\
\end{tabular}
   \caption{Summary of our results for the CR and BR models on reproducing the claimed echoes following GW170817. Integers $m$ and $n$ are the azimuthal number and overtone indices of the ringdown QNM that dominate echoes, while $l$ is fixed to 2. }
   \label{tab:chart}
\end{table}

\subsection{Failed supernovae}
The death of massive stars and their gravitational collapse can lead to very diverse outcomes. Stellar evolution calculations predict that massive stars within a certain mass range have compact cores which do not explode as supernovae and instead directly collapse into BHs (e.g. \cite{OConnor:2010moj, Sukhbold:2015wba}). The existence of such {\it failed supernovae} is suggested as a solution to the discrepancy between the cosmic supernova and star-formation rates \citep{Horiuchi:2011zz}, and the absence of red supergiant stars of mass $\gtrsim 17\ {\rm M_\odot}$ as supernova progenitors \citep{Smartt:2008zd,Horiuchi:2014ska,Kochanek:2014mwa}. A survey monitoring nearby evolved massive stars has been carried out to directly find these failed supernovae \citep{Kochanek:2008mp,Gerke:2014ooa}, and recently found a strong candidate of a BH forming from a vanishing red supergiant star \citep{Gerke:2014ooa,Adams:2016ffj}. In this section, we consider echoes from BHs which form from collapses of nearby massive stars.

Several numerical studies have estimated the GW emission from collapse of NSs \cite{PhysRevLett.55.891,Sekiguchi:2005bi, Baiotti:2007np,Ott:2010gv,Giacomazzo:2011cv,Cerda-Duran:2013swa}. In particular, \cite{Giacomazzo:2011cv} obtained a fit to numerical simulations that predicts GW emission from BH formation over a wide range of spin with $\bar{a} \gtrsim 0.1$ \cite{Baiotti:2007np}: 
\begin{align}
\epsilon_{\text{rd}} \approx \frac{\bar{a}^{n_1}}{C_1\bar{a}^{n_2}+C_2},
\label{eq:epsilon_rd_failedSN}
\end{align}
where the fitting parameters are $n_1=1.43\pm 0.74, n_2=2.63\pm 0.53, C_1=(5.17\pm4.37)\times 10^5$, and $C_2=(1.11\pm 0.57)\times 10^6$. The value of $\epsilon_{\text{rd}}$ is typically of order $10^{-8}$--$10^{-6}$, being larger for higher BH spin. 

Before doing a detailed calculation of echoes, we first do a back-of-the-envelope estimate of its detectability using the results for BNSs in the previous section. For the GW170817 case, the energy emitted as GW echoes within 500 Hz is obtained as $\sim 8\times 10^{-4}M_\odot c^2$ \cite{Abedi:2018npz}, which is $\sim 3$\% of our assumed ringdown energy $\epsilon_{\rm rd}Mc^2\sim 2.7\times 10^{-2}M_\odot c^2$, where we used $\epsilon_{\rm rd} = 1$\% from the above fit. Assuming the fraction is preserved for this case, and the echoes is emitted by a BH of $M\sim 2.5M_\odot$ at frequency $f_{\rm echo} \sim 100$ Hz, the strain at distance $D$ is roughly
\begin{align}
h &\sim \sqrt{\frac{(0.03\epsilon_{\rm rd}Mc^2) G}{\pi^2 c^3D^2f_{\rm echo}^2}} \nonumber \\
&\sim 4\times 10^{-25}\ {\rm Hz}^{-1/2} \nonumber \\
&\times \left(\frac{\epsilon_{\rm rd}}{10^{-7}}\right)^{1/2}\left(\frac{f_{\rm echo}}{100{\rm\  Hz}}\right)^{-1}\left(\frac{D}{10{\rm\ Mpc}}\right)^{-1}.
\end{align}
Thus current 
(or 2nd generation)  detectors at design sensitivity with noise power spectrum $\sqrt{S_n}\sim 10^{-24}\ {\rm Hz}^{-1/2}$ can detect sources out to $\sim1$ Mpc, while 3rd generations detectors with $\sqrt{S_n}\sim 10^{-25}\ {\rm Hz}^{-1/2}$ may reach out to $\sim 10$ Mpc.

\begin{figure}[t]
    \includegraphics[width=0.45\textwidth]{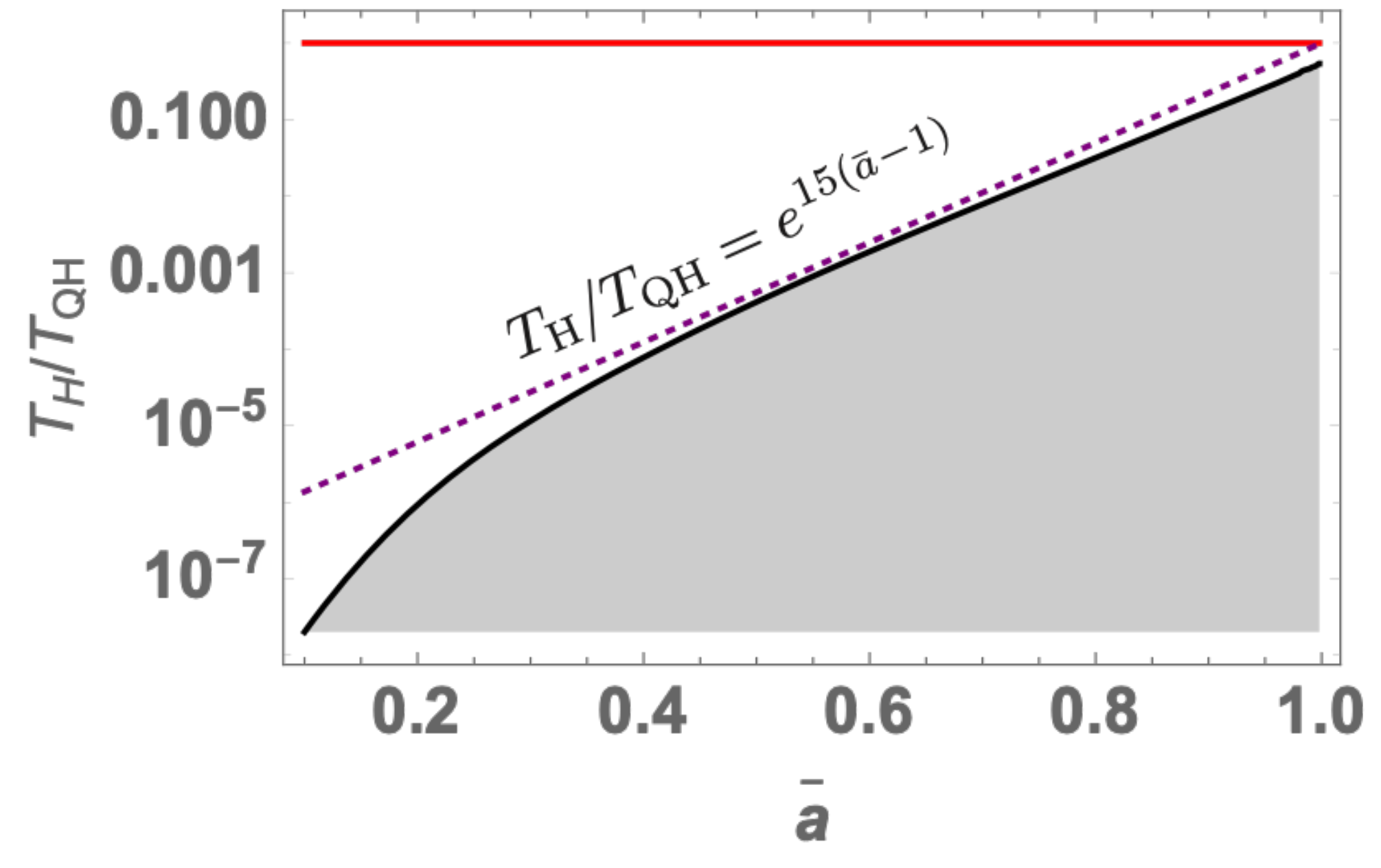}
\caption{Red solid and purple dashed lines show the value of $T_{\rm H}/T_{\rm QH} (\bar{a})$ we use to calculate the horizon distances. The former one is $T_{\rm H} / T_{\rm QH} = 1$ and the latter one is an optimistic case in which the reflection rate is enhanced while avoiding the ergoregion instability (gray region) in the range of $0\leq \bar{a} \leq 0.998$.
}
\label{supernova_alpha}
\end{figure}

To calculate the detailed spectra and detectable distance, we consider BH masses and spins in the range $2.1 M_\odot \leq M \leq 3 M_\odot$ and $0.1 \leq \bar{a} \leq 0.9$. We assume the GWs are emitted as ringdown radiation just after the collapse, and is dominant in the $l=2, m=0$ mode, which is justified if the system does not deviate significantly from axisymmetry. We thus use the value of $\epsilon_{\rm rd}$ in equation \eqref{eq:epsilon_rd_failedSN} to obtain the QNM amplitude for $l=2$ and $m=0$. The outgoing echo spectrum $h_{\rm echo}(f)$ is calculated using the BR (i-iii) and CR (iv) models for the following cases:
\begin{enumerate}
    \item An $n=0$ (fundamental) mode with $T_{\rm H}/T_{\rm QH}=1$,
    \item $n=0$ mode with $T_{\rm H}/T_{\rm QH} = e^{15 (\bar{a}-1)}$ (see Fig. \ref{supernova_alpha}),
    \item An $n=2$ overtone with $T_{\rm H}/T_{\rm QH}=0.54$,
    \item An $n=2$ overtone with $R_c = 0.71$.
\end{enumerate}
The last three cases are what we learned in the previous section to be consistent with the claimed echoes following GW170817. For the first two cases we adopt $\gamma = 10^{-10}$, and for the last two cases $\gamma=1$. We also assume ${\cal F} = 1$ in this subsection, i.e. the source is at an optimal sky location.

We calculate the optimal signal to noise ratio $\rho_{\rm opt}$ from
\begin{equation}
\rho_{\rm opt}^2 = \int \frac{\tilde{h}^2(f)}{S_n(f)} df,
\end{equation}
where $\tilde{h}(f)$ is the calculated echo signal, and $S_n(f)$ is the noise spectral density. The viewing angle is assumed to be orthogonal to the BH spin, which is the optimal case for $l=2, m=0$ modes (see Fig 5-(a)). To obtain $S_n(f)$ we use the expected sensitivity curves for Advanced LIGO at Design Sensitivity \cite{Evans:2016mbw} and the two 3rd generation detectors, Einstein Telescope \cite{Hild:2010id} and Cosmic Explorer \cite{Evans:2016mbw}.

Using the signal to noise ratio we calculate the horizon distance, defined here as the distance where $\rho_{\rm opt}=8$. We note that this is a definition widely used in the context for searching inspirals of compact object binaries (e.g. \cite{OShaughnessy:2009szr}), and is valid only if we have an accurate template waveform. If the true reflectivity near the horizon is different from the BR and CR models assumed here, the signal would not match the theoretical waveform (although resonance frequencies are likely unchanged), thus reducing the distance out to which echoes are observable.

\begin{figure*}
    \includegraphics[width=0.8\textwidth]{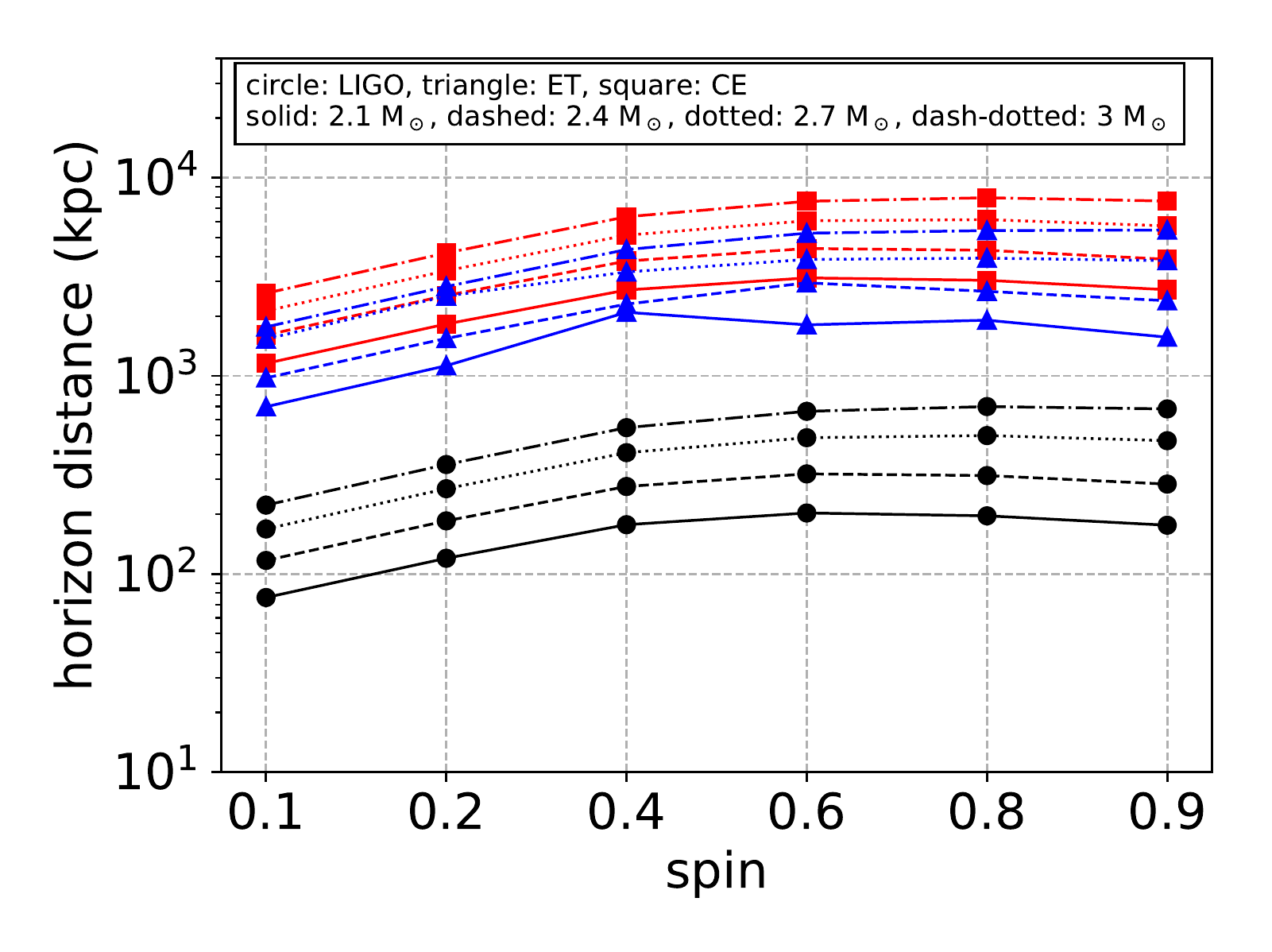}
\caption{Horizon distances for various BH masses and spins for case (i) in main text: the BR case with $T_{\rm H}/T_{\rm QH} = 1$, and $n=0$ (fundamental) mode is dominant. The blue dots, orange triangles and green squares are obtained respectively by using the expected noise curves for the LIGO Design sensitivity, Einstein Telescope, and Cosmic Explorer.
}
\label{fig:horizon_case1}
\end{figure*}

\begin{figure*}
    \includegraphics[width=0.8\textwidth]{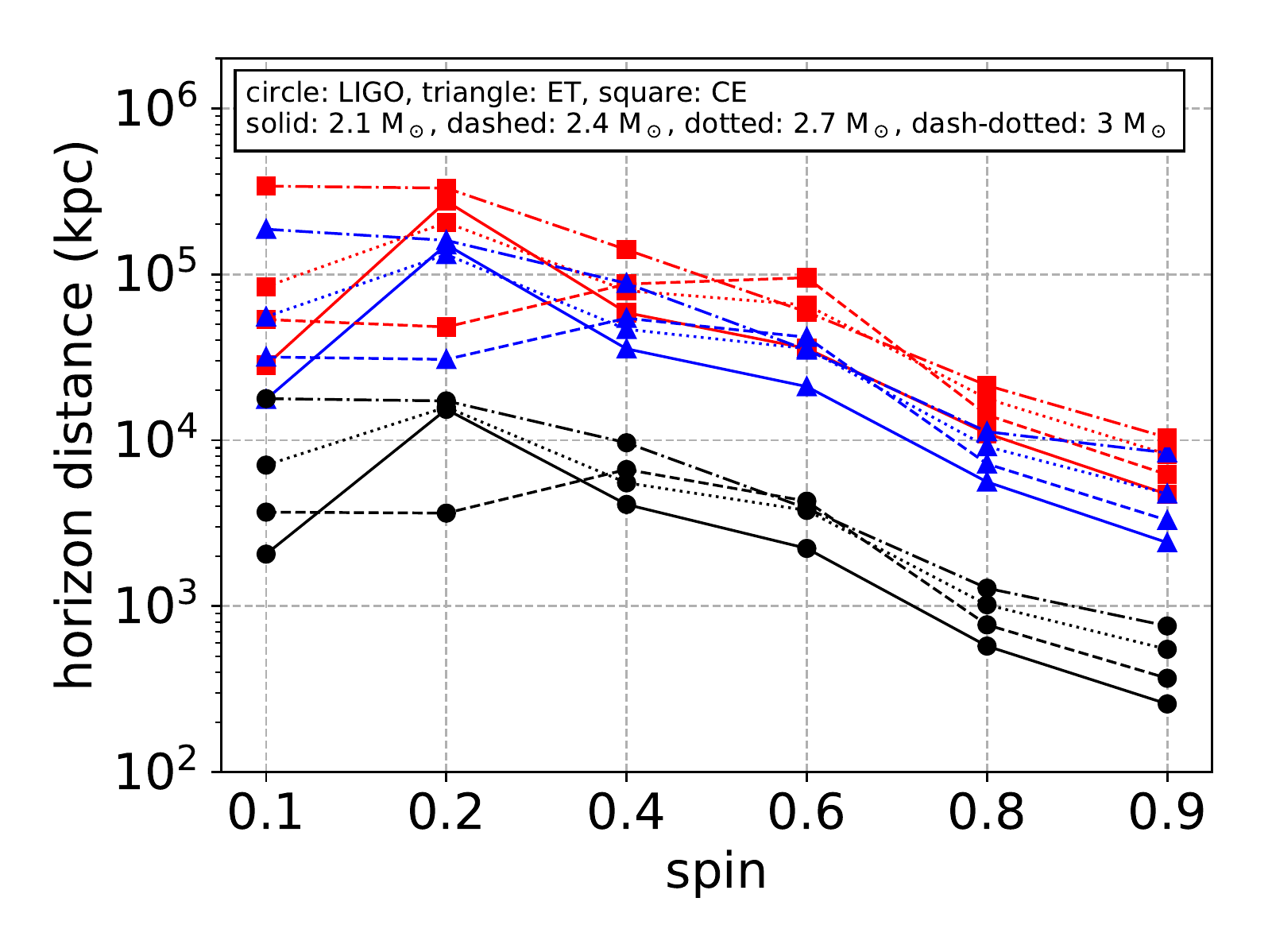}
\caption{Same as Fig \ref{fig:horizon_case1}, but horizon distances for case (ii).
}
\label{fig:horizon_case2}
\end{figure*}

\begin{figure*}
    \includegraphics[width=0.8\textwidth]{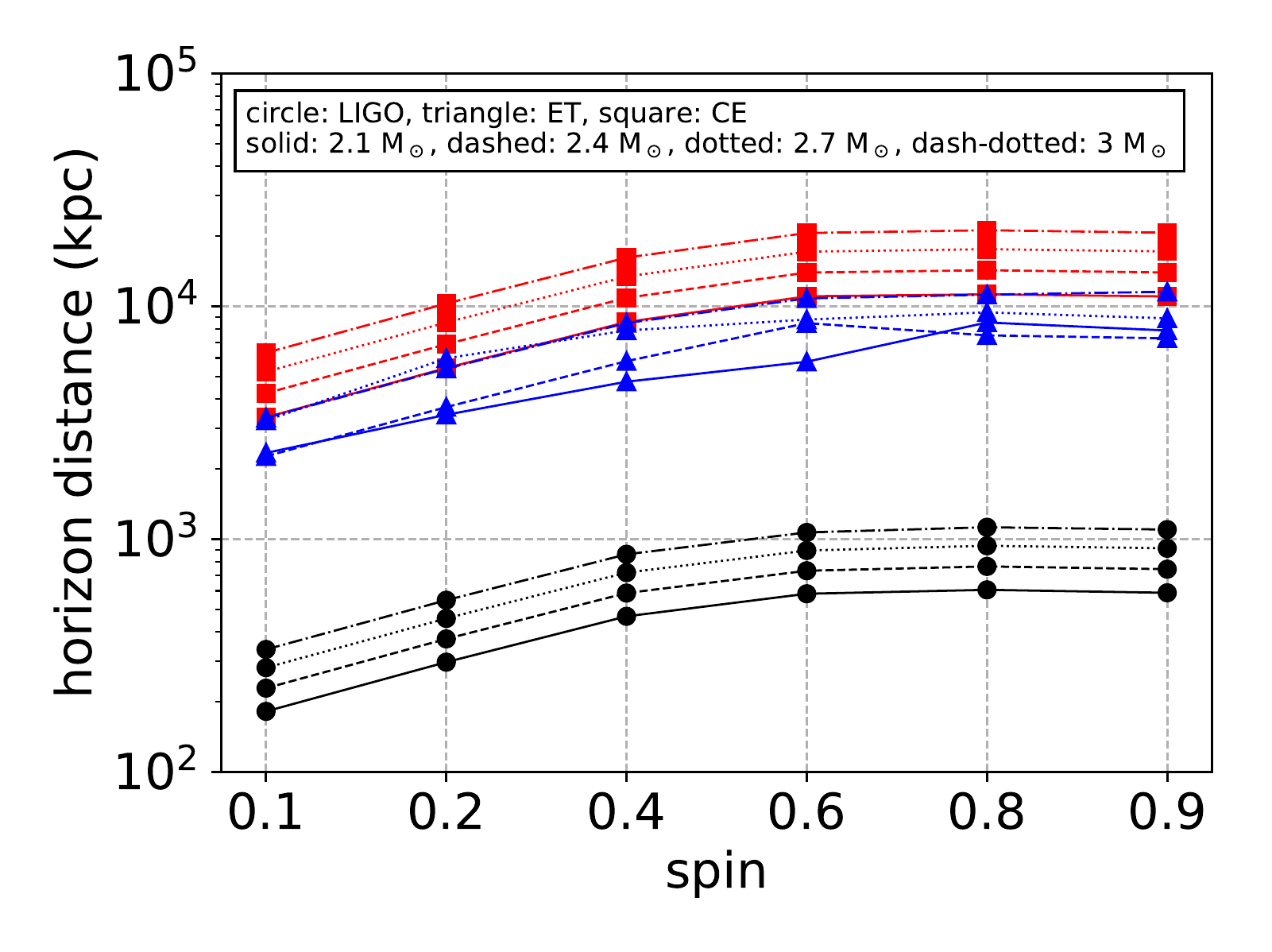}
\caption{Same as Fig \ref{fig:horizon_case1}, but horizon distances for case (iii).
}
\label{fig:horizon_case3}
\end{figure*}

\begin{figure*}
    \includegraphics[width=0.8\textwidth]{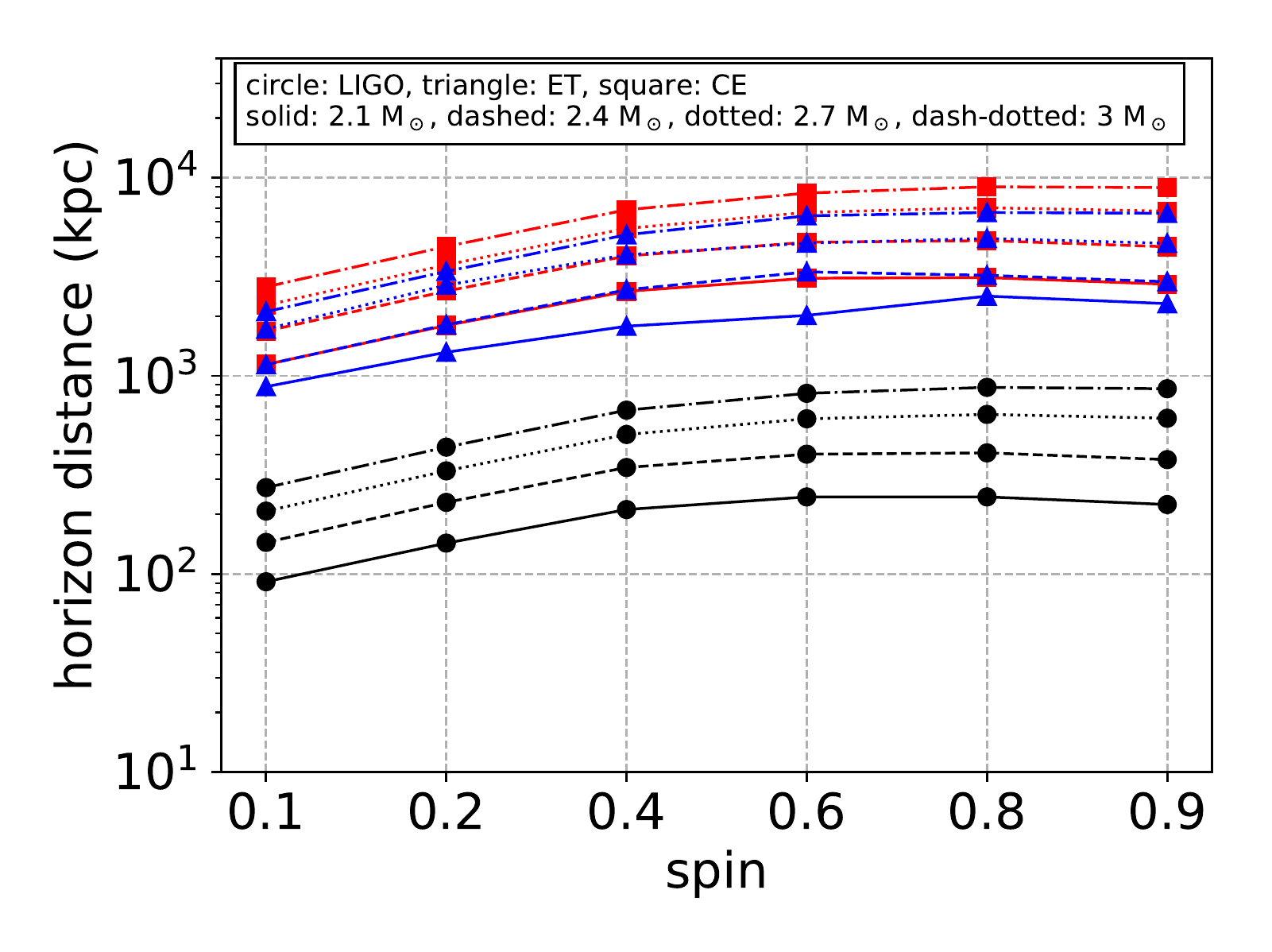}
\caption{Same as Fig \ref{fig:horizon_case1}, but horizon distances for case (iv).
}
\label{fig:horizon_case4}
\end{figure*}

Fig. \ref{fig:horizon_case1} to \ref{fig:horizon_case4} show the horizon distances for the three cases assumed above. For cases (i) and (iii) of $T_{\rm H}/T_{\rm QH}=1$, we generally see an increase in the horizon distance for higher spins, due to the spin dependence of $\epsilon_{\rm rd}$. However for the case (ii), the exponential dependence of the reflectivity on the BH spin gives a larger horizon distance for smaller spins.

For case (i), the horizon distance lies at $0.1$ -- $0.5$ Mpc for Advanced LIGO, and at $1$--$8$ Mpc for the third-generation detectors depending on spin. For the other three cases, the range extends out to $1$ Mpc for Advanced LIGO, and a few 10 Mpc for the third-generation detectors depending on spin. We can thus conclude that echoes can be probed in the future for failed supernovae that occur in the nearby universe. 

\begin{figure}[t]
    \includegraphics[width=0.45\textwidth]{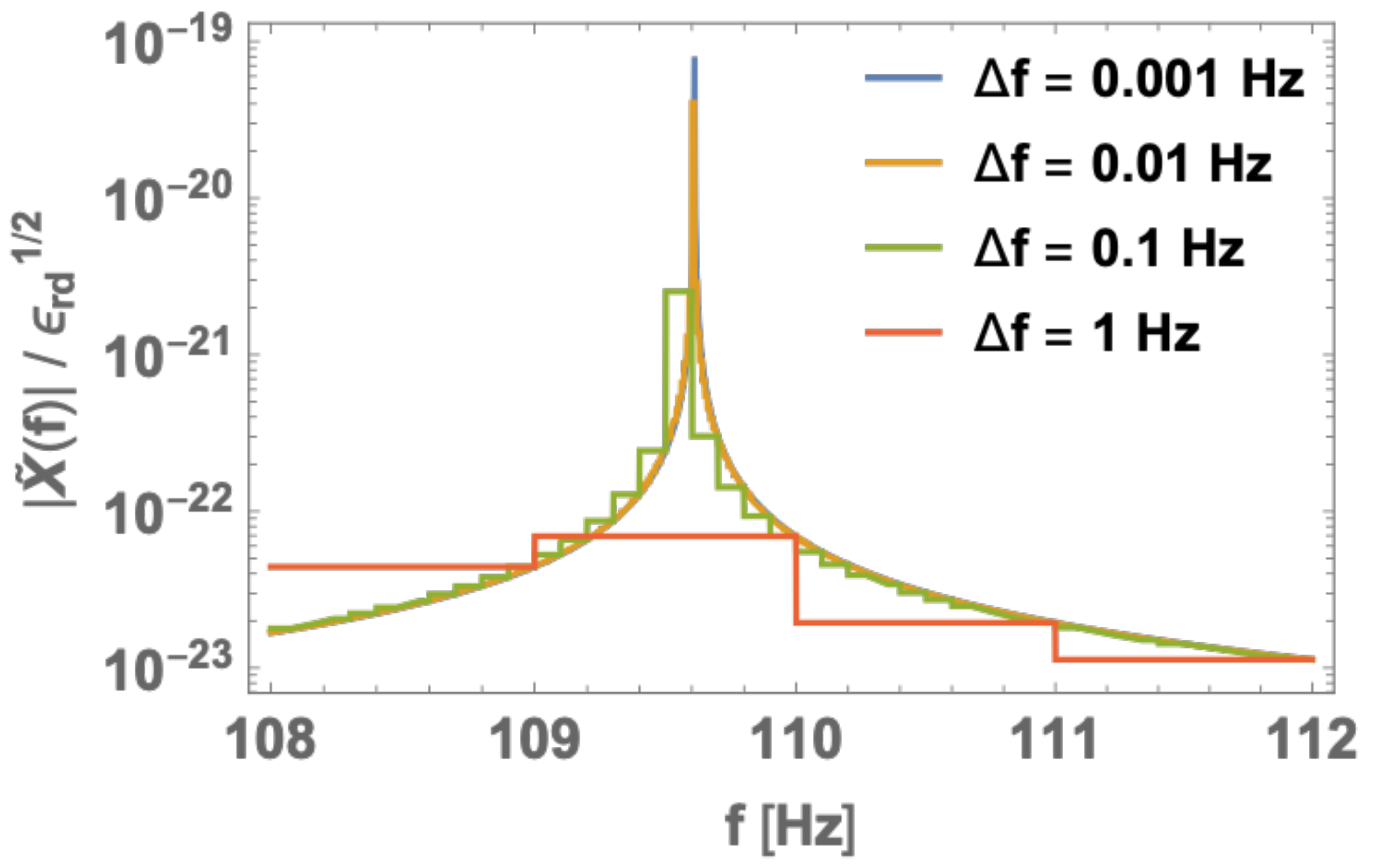}
\caption{Spectra of GW echoes in the BR model in $108 - 112$ Hz with $\ell = 2$, $m=0$, $\bar{a} = 0.1$, $M = 2.1 M_{\odot}$, $D_L = 40$ Mpc, $T_{\rm H} / T_{\rm QH} = e^{15 (\bar{a} -1)}$, and $\gamma = 10^{-10}$. The frequency resolution $\Delta f$ is set to $1$ Hz (red), $0.1$ Hz (green), $0.01$ Hz (yellow), and $0.001$ Hz (blue).
}
\label{resolution}
\end{figure}

We however note that the detectability is affected by how long the waveform template is, especially for case (ii) where $T_{\rm H}/T_{\rm QH}$ can become very low and create a very long-lived signal. In the above estimates we used a spectral resolution of 0.1 Hz for $\tilde{h}(f)$, which corresponds to using a template of $10$ seconds long for matched filtering. We plot the spectrum for different spectral resolution, with all the model parameters fixed (Fig. \ref{resolution}). We find that a coarser spectral resolution reduces the SNR and thus horizon distance by almost an order of magnitude for low spins. However, note that this is an artefact of the choice $T_{\rm H} / T_{\rm QH} = e^{15 (\bar{a} -1)}$, marking the boundary of ergoregion instability in Fig. \ref{supernova_alpha}. For values of $T_{\rm QH}$ far from this boundary, e.g., cases (i) and (iii) where $T_H\sim T_{\rm QH}$, the damping time is short enough that template duration does not severely affect the results.

Recently, a search for failed supernovae monitoring evolved massive stars in nearby galaxies found a strong candidate at a distance of $6$ Mpc \cite{Adams:2016ffj}. If candidates like this are found in the future, it may be of interest to conduct follow-up search for echoes. Assuming that BHs form in $\eta \sim 15 \%$ of core-collapse \cite{Adams:2016hit}, the rate $r$ of BH formation within distance $d$ is
\begin{eqnarray}
r \sim 0.4\ {\rm yr}^{-1} \left(\frac{\eta/(1-\eta)}{0.18}\right)\left(\frac{d}{20 \ {\rm Mpc}}\right)^{-3}
\end{eqnarray}
where we used the local (successful) core-collapse supernova rate $\sim 7\times 10^{-5}\ {\rm Mpc}^{-3}\  {\rm yr}^{-1}$ \cite{Li_2011}. Within these horizon distances, electromagnetic emissions from failed supernovae may be detectable (e.g. \cite{Piro:2013voa,Lovegrove:2013hox,Fernandez:2017pcd,Tsuna:failedSN}). A multi-messenger search can help (i) significantly narrow down the time window to search for a possible echo signal, and (ii) decrease the detection threshold, if a viable candidate is found. 

Since echoes can be seen when the remnant NS reaches its maximum mass and collapses to a BH, extraction of the peak frequency may help constrain the NS's maximum mass. The dependence of $\Delta t_{\rm echo}$ on BH spin (equation \ref{eq:fecho}) implies that BHs formed at lower spins are better targets for pinning down this mass. We note however that thermal effects can significantly change the maximum mass (by $\mathcal{O}(10\%)$ \cite{Schneider:2020kxr}), and the maximum mass for a cold NS one derives will depend on the model one assumes for this effect. Although detailed discussion on the feasibility of extracting the BH parameters is beyond the scope of this work, echoes may provide crucial and complementary information for constraining the equation of state of NSs.


\section{Conclusion}
In Paper I, we provided a formulation for quantum black hole seismology, which could be used to calculate spectra of gravitational wave echoes emitted by spinning black holes. Utilizing this framework, in this work we considered the echo signatures from two representative astrophysical events leading to black hole formation: neutron star mergers and failed supernovae. 

For neutron star mergers, we compared the Boltzmann and constant reflectivity models against the detection claim of gravitational wave echoes from GW170817. We find that if the echoes are energetically dominated by the fundamental quasinormal modes of the ringdown, both models have difficulty reproducing the claimed strength of echoes. This is because the required reflectivity is too high and necessitates an unnatural spin-dependence of the reflectivity in order to avoid the ergoregion instability. They also require super-Planckian frequencies near the horizons of the black hole. 

However if the overtone quasinormal modes of $\ell = 2$, $m=0$ ($\ell = m = 2$) are highly excited when the ringdown phase starts, the echo peaks at low frequencies are strongly enhanced in the Boltzmann (constant) reflectivity model, consistent with the level claimed for GW170817 \cite{Abedi:2018npz}. However for the case of the Boltzmann reflectivity model with $m=0$ mode, a larger ringdown energy $\epsilon_{\rm rd}$ than predicted from simulations or a spin-dependence of reflectivity is required. Although the Boltzmann reflectivity may be more physically motivated, one may favor the constant reflectivity model between the two.

For failed supernovae, we calculated the maximum distance to which we can observe the echoes predicted by the Boltzmann reflectivity model. We find that both second-generation and third-generation detectors may see (or constrain) echoes of black holes in the nearby universe, out to a few $\times 10$ Mpc for third-generation detectors. We claim that failed supernovae can be an additional interesting probe of quantum gravity, as well as a potential probe of the equation of state for neutron stars.

Our echo calculation assumes a simple ringdown waveform of a single quasinormal mode for the input. This is obviously an approximation, as works utilizing numerical relativity simulations find more complex waveforms post-merger. Using more realistic waveforms from these simulations would be an important improvement to our analysis.

We shall further note that although the evidence claimed for gravitational wave echoes in compact binary mergers are still controversial, they have significantly driven the scientific community toward constructing a theoretical framework for phenomenology of quantum black holes and exotic compact objects \cite{Abedi:2020ujo}. As theoretical modeling has improved, it is now time to infer the correct model from observations of black hole-forming events. This is very hopeful in the foreseeable future, as detectors with enough sensitivity to probe this exotic phenomenon are starting to be realized.

\begin{acknowledgements}
We thank Edoardo Milotti for valuable comments that improved this manuscript. We also thank Kazumi Kashiyama and Daniel Siegel for helpful comments and discussions. We also thank all the participants in our weekly group meetings for their patience during our discussions. D. T. thanks the Perimeter Institute for the hospitality during his visit. This work was supported by the University of Waterloo, Natural Sciences and Engineering Research Council of Canada (NSERC), and the Perimeter Institute for Theoretical Physics. N. O. is supported by the JSPS Overseas Research Fellowships. D.T. is supported by the Advanced Leading Graduate Course for Photon Science (ALPS) at the University of Tokyo, and by JSPS KAKENHI grant No JP19J21578. Research at Perimeter Institute is supported in part by the Government of Canada through the Department of Innovation, Science and Economic Development Canada and by the Province of Ontario through the Ministry of Colleges and Universities.
\end{acknowledgements}

\appendix
\section{calculation of $h_{H,L}$}
\label{app:h}
In this appendix, we provide our methodology to include echo signal into the detected GW waveforms in Hanford and Livingston. The observed waveform is
\begin{equation}
h_{H,L}(t) = F_+ h_+(t) + F_{\times} h_{\times} (t),
\label{app_wave}
\end{equation}
and the spectrum of $h_{+, \times}$ is obtained as
\begin{equation}
\tilde{h}_{+, \times} (\omega) = \tilde{h}_{+, \times}^{(\text{rd})} (\omega) (1+ {\cal K} (\omega)),
\label{app_spect}
\end{equation}
where ${\cal K}$ is the transfer function introduced in the Paper I. Substituting (\ref{app_spect}) into the Fourier form of (\ref{app_wave}), the spectrum is
\begin{align}
\begin{split}
\tilde{h}_{H,L} (\omega) &= (F_+ \tilde{h}_{+}^{(\text{rd})} + F_{\times} \tilde{h}_{\times}^{(\text{rd})}) (1+ {\cal K})\\
&= \tilde{h}_{H,L}^{(\text{rd})} (\omega) (1+ {\cal K}),
\end{split}
\end{align}
and $\tilde{h}_{H,L}^{(\text{rd})}$ is given in (\ref{spectrum_rd}). We then substitute $\tilde{h}_{H,L}$ into (\ref{eq:spec_H}) and (\ref{eq:spec_L}) with the proper normalization to obtain the functions $H(t,f)$ and $L(t,f)$.

\bibliography{reference}

\end{document}